# Impact of repetitive, ultra-short soft X-ray pulses from processing of steel with ultrafast lasers on human cell cultures

Julian Holland#,*, , Cristiana Lungu#, ✉, **, *** , Rudolf Weber✉,* , Max Emperle**** and Thomas Graf*

*Institut für Strahlwerkzeuge, University of Stuttgart, Pfaffenwaldring 43, 70569 Stuttgart, Germany
**Institute of Cell Biology and Immunology, University of Stuttgart, Allmandring 31, 70569, Stuttgart, Germany
***Stuttgart Research Center Systems Biology (SRCSB), University of Stuttgart, Nobelstraße 15, 70569, Stuttgart, Germany
****Institute for Biochemistry and Technical Biochemistry, University of Stuttgart, Allmandring 31, 70569, Stuttgart, Germany

# These authors have contributed equally to this work
✉ Correspondence to: rudolf.weber@ifsw.uni-stuttgart.de or cristiana.lungu@izi.uni-stuttgart.de

**Abstract**

Ultrafast lasers, with pulse durations below a few picoseconds, are of significant interest to the industry, offering a cutting-edge approach to enhancing manufacturing processes and enabling the fabrication of intricate components with unparalleled accuracy. When processing metals at irradiances exceeding the evaporation threshold of about $10^{10}$ W/cm² these processes can generate ultra-short, soft X-ray pulses with photon energies above 5 keV. This has prompted extensive discussions and regulatory measures on radiation safety. However, the impact of these ultra-short X-ray pulses on molecular pathways in the context of living cells, has not been investigated so far.
Using laser pulses of 250 fs and 6.7 ps, along with pulse repetition rates exceeding 10 kHz, we conducted the first molecular characterization of epithelial cell responses to ultra-short soft X-ray pulses generated during processing of steel with ultrafast lasers. Ambient exposure of vitro human cell cultures, followed by imaging of the DNA damage response and fitting of the data to an experimentally calibrated model of dose rate estimation, revealed a linear increase in the DNA damage response relative to the exposure dose. This is in line with findings from work using continuous wave soft X-ray sources and suggests that the ultra-short X-ray pulses do not generate additional hazard. This research contributes valuable insights into the biological effects of ultrafast laser processes and their potential implications for user safety.

**Introduction**

Ultra-short laser pulses with pulse durations below a few picoseconds and pulse repetition rates up to several GHz allow processing of almost any material with very high precision and accuracy well below 1 μm[1]. Recently this technology gained importance for industrial applications due to the rapid increase of the pulse energy, the frequency and in particular of the average output power of the so-called ultrafast lasers. Industrial processes include e.g., micro-cutting, surface structuring and drilling of metals, where a significant part of the material is evaporated. The threshold-irradiance for evaporation of metals is a few times $10^{10}$ W/cm², the optimum process window regarding efficiency and quality is in the range of $10^{11}$ W/cm² to $10^{13}$ W/cm² and occurs at ambient pressure[2]. However, with such lasers irradiances even exceeding $10^{15}$ W/cm² can be generated[1]. Above the evaporation threshold the laser pulses produce a plasma, which emits radiation that has a measurable part in the soft X-ray region with photon energies above 5 keV for irradiances exceeding about $10^{13}$ W/cm$^2$. It is important to note that each laser pulse creates a soft X-ray pulse, which has a similar duration as the laser pulse. Hence, also the X-ray pulses are ultra-short and have high peak powers. At distances closer than about 30 cm to the plasma, the soft X-ray spectrum features a peak at a photon energy near 5 keV[3–6], which results from both, the plasma emission spectrum and the absorption in air. These soft X-rays were subject to investigations in several recently published articles with the emphasis on the resulting dose rates depending on processing strategies,[7] laser parameters,[7–10] and processed materials[11]. The major issue of the emitted soft X-rays is the potential hazard for the operator. Several publications address the hazardous potential[6,12] and the safety aspects[5,13] arising from this radiation. In Legall et. al.[5] and Weber et. al.[3] it was measured that when increasing the irradiance from about $10^{13}$ W/cm$^2$ by a factor of 10, the dose rate per Watt of average laser power increases by about a factor of 10 000. This highlights the critical relationship between irradiance and laser pulse duration: as pulse duration decreases, irradiance increases, assuming all other processing parameters remain unchanged. On the other hand, research by Holland et. al.[10] found that, when keeping the irradiance constant, the dose rate decreases by almost a factor 1 000 when decreasing the pulse duration from 5 ps to 75 fs. Furthermore, the impact that these very short soft X-ray pulses resulting from industrial materials processing with ultrafast lasers have on biological systems, specifically at the cellular level, is not known. Therefore, it is not clear, if shortest laser pulses might have an increased potential for hazard.

The absorption of X-ray radiation in the tissue of the operator has the potential to induce molecular damage, thereby posing a risk to human health. It is widely established[14–17] that most of the biologically important damage caused by X-ray exposure is the one to the DNA[18]. This damage can be either direct, if photons and secondary electrons hit the DNA directly, or indirect, as a consequence of free radicals that are produced in the vicinity. These insults can result in a broad pattern of lesions, among which double stranded breaks (DSBs) are considered the most dangerous. Erroneous or incomplete DSB repair can lead to mutations or large chromosomal rearrangements that can trigger diverse responses such as cell death or transformation. The phosphorylation of histone H2AX on serine 139 (γH2AX) is one of the earliest signals in the pathway of the DNA damage response and accumulates near DSBs. This modification has been reliably used as an in vitro and in vivo marker to measure double strand DNA damage induced by X-ray exposure [19,20].

The aim of the present study was to gain a first molecular quantification of possible hazard of the soft X-radiation caused during industrial materials processing with ultrafast lasers. Therefore, the cervical adenocarcinoma epithelial cell line HeLa was used as a model system to monitor in vitro the cellular response to exposure to soft X-radiation created during processing of steel with ultra-short laser pulses. The resulting DSBs response was quantified by microscopy on the basis of the γH2AX signal.

**Materials and methods**

Generation of soft X-rays

Two different setups were used for the generation of the soft X-rays used for exposing the cells. On the one hand, a setup with a very short laser pulse duration of 250 fs and moderate average power of up to 14 W was used. The Pharos laser system (Light Conversion, Vilnius, Lithuania) used in this setup had a laser wavelength of 1030 nm, a pulse repetition rate of 50 kHz, and a maximum pulse energy of 280 µJ. The beam was focused with an F-theta lens to a spot with a diameter of 57 µm on the surface of a 1.4301-steel sample (72% Fe, 18% Cr, 9% Ni, <1% Mn,Si). The resulting maximum irradiance on the sample was $8.7 \cdot 10^{13}$ W/cm². On the other hand, a setup with a comparatively long laser pulse duration of 6.7 ps and a very high average power of up to 500 W was used. The custom laser system consisting of a TruMicro5000 (Trumpf, Ditzingen, Germany) seed laser and an IFSW multi-pass amplifier[21] used in this setup had a laser wavelength of 1030 nm, a pulse repetition rate of 300 kHz, and a maximum pulse energy of 1.66 mJ. The same focusing optics was used as described above. The resulting focus diameter was 46 µm, yielding a maximum irradiance of $2.5 \cdot 10^{13}$ W/cm² on the steel sample. The arrangement of the experiments is shown in Figure 1. The table in the inset on the right side summarizes the distances and angles of exposure relative to the normal of the surface of the dishes.

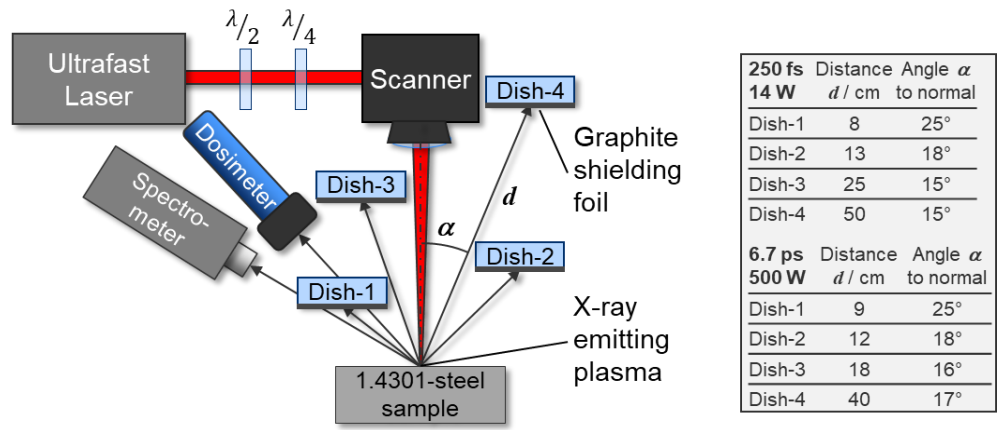

*Figure 1. Experimental setup. The beam was moved with a scanner and focused with an F-Theta lens onto a 1.4301-stainless-steel sample. Four dishes with cells were positioned at four distances ($d_1$-$d_4$) and were exposed to the X-rays at the same time. The bottom of the dishes was covered with a 200 µm thick carbon foil to block visible and UV-radiation. The dose rates were measured with an OD-02 at different distances $d_{OD}$ from the processing zone, and the X-ray spectrum was measured at a distance of $d_{PN}$. The table in the inset summarizes the distances and angles of exposure relative to the normal of the surface of the dishes.*

A square area of 2.5 x 2.5 cm² of the surface of the sample was processed by applying a hatch pattern with parallel lines. The pulse-to-pulse overlap was 80% and the line-to-line overlap 70% of the beam diameter on the sample surface. The scanning direction was changed by 45° after each processed plane to avoid the formation of grooves, which can cause shielding of the X-ray emission. In addition, these scanning parameters were consistent with those proposed by Legall et. al[7] to optimize the emission yield. The resulting $\dot{H}(0.07)$ dose rates were measured in mSv/h with an OD-02 detector (STEP, Pockau-Lengefeld, Germany) at the distance $d_{OD}$ to the processing region.

Spectrum of the soft X-rays

The spectrum of the X-ray emission from plasmas, which are produced with ultra-short laser pulses consists of line emission, recombination radiation and Bremsstrahlung. However, the emission of the plasmas considered in this work is dominated by Bremsstrahlung[4]. This part of the radiation can be described by a model introduced by Weber et. al.[3,22] which predicts the resulting dose rates of the X-ray emission from laser-produced plasmas. Figure 2 shows a typical spectrum (black) measured with a spectrometer (XRS-30-128-100-BeP Complete, PN-Detector, Munich, Germany) at a distance of $d_{PN}$ = 25 cm while processing stainless steel (1.4301) with an irradiance of $8.7 \cdot 10^{13}$ W/cm². The calculation of the Bremsstrahlung with the model is shown as a red line and is in very good agreement with the measurement. The measured spectrum also shows the line emission of iron and the alloying elements Mn, Si, and Ni. However, line emission contributes only a few percent to the total X-ray emission at the current process parameters[4] and can therefore be neglected for the calculation of the resulting dose rate.

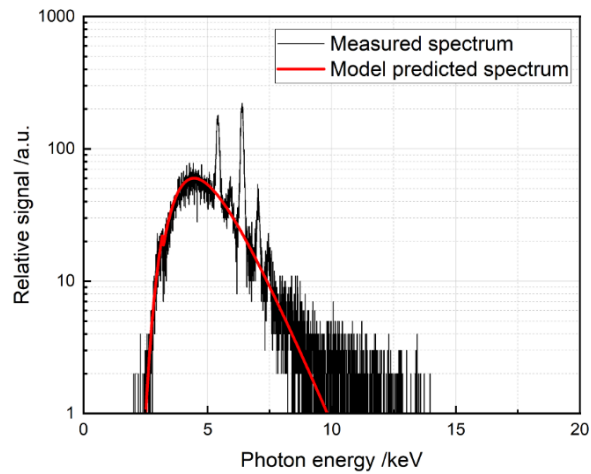

*Figure 2. Typical X-ray emission spectrum from the plasma produced during ultrafast laser processing of stainless steel with an irradiance of 8.7·10$^{13}$ W/cm² as detected at a distance of 25 cm from the processing zone. The major part of the emission consists of Bremsstrahlung emission with a maximum at the photon energy of about 4 keV at this distance. The red line refers to the model calculation of the expected Bremsstrahlung radiation.*

Determination of the soft X-ray transmission of the dish

The $\dot{H}'(0.07)$ dose rates were measured with and without an empty cell culture dish made of polystyrene (PS) in the line of sight between OD-02 and the X-ray emitting plasma. The measured values were compared with the model predictions, considering the thickness of the bottom of the dish of 0.92 mm and the PS material. The OD-02 was placed at a distance of 18±2 cm from the processing zone and the calculated laser irradiance at the surface of the steel sample was 3.4·10$^{13}$ W/cm². The maximum measured dose rate without a dish was 519 mSv/h, which is in very good agreement with the calculation of the model, which predicts a maximum $\dot{H}'(0.07)$ of 522 mSv/h.
With the PS-dish between the plasma and the OD-02, the measured dose rate decreased by approximately 60% to 203 mSv/h. For the model predictions with the dish as a filter material, the molar mass of the PS was varied for the calculation of the transmission coefficients[23] until the calculation was in a good agreement with the measurement. The molar mass for the best fit of the model to the measurement was found to be $M_{PS}$ = 37 amu. Furthermore, the density assumed in the calculation of the transmission coefficient of PS was 0.91 gcm$^{-3}$. The dose rate calculated with the model including the parameters for the PS-dish was 202 mSv/h.

Cell culture

HeLa cells were obtained from ATCC and were cultured in RPMI 1640 (Invitrogen) supplemented with 10% foetal calf serum (FCS; PAA Laboratories). The cells were maintained in a humidified incubator at 37°C with 5% $CO_2$. The cell line was authenticated, tested negative for Mycoplasma (Lonza, LT07-318) and kept in culture for no longer than one month.

Exposure of HeLa cells to X-rays from laser produced plasma

3 x 10$^5$ HeLa cells were seeded on 35 x10 mm$^2$ PS-dishes (Greiner, 627160) and were exposed 48 h later to X-rays from laser produced plasma. The exposure was performed in air and at room temperature, using the two aforementioned laser systems with largely different processing parameters, according to the scheme shown in Figure 1. For each experiment, four dishes were positioned inside the enclosure at different distances relative to the processing region, labelled with Dish-1 to Dish-4. In addition, one dish was placed outside of the laser enclosure and served as the non-exposed control. The distances of the dishes were varied between the experiments for practical reasons, and the exact values are given in the results section. Before exposure, the bottom of the cell dishes was covered with a 200 µm thick carbon foil (C) to block the visible and UV-radiation radiation <100 eV, which is also emitted from the laser-induced plasma[23].

Determination of the effective dose absorbed by the cells

At the time of X-ray exposure, the PS-dishes were covered with a cell monolayer of ca. 80% confluency. Based on literature, the height of the epithelial monolayer was approximated to 6 µm[24]. Therefore, the $\dot{H}(0.005)$-dose rate was chosen to quantify the absorbed dose. As the OD-02 dosimeter only measures the standard $\dot{H}(0.07)$-dose rates, the model described by Weber[3,10] was used to determine the $\dot{H}(0.005)$-dose rates. In a first step, the model was validated by comparing the values calculated with the model (Figure 3, dash-dotted lines), which includes both, the transmission of the 200 µm-thick graphite foil and of the bottom of the dish, with the values measured with the OD-02 dosimeter (diamonds). The error bars are the standard deviation of the dose rate measurements. The identical calibration of the model was used for the experiments performed with both experimental setups. In the second step, the model was used to calculate the $\dot{H}(0.005)$-dose rates (Figure 3, solid lines) to which the cells in the monolayer were exposed.

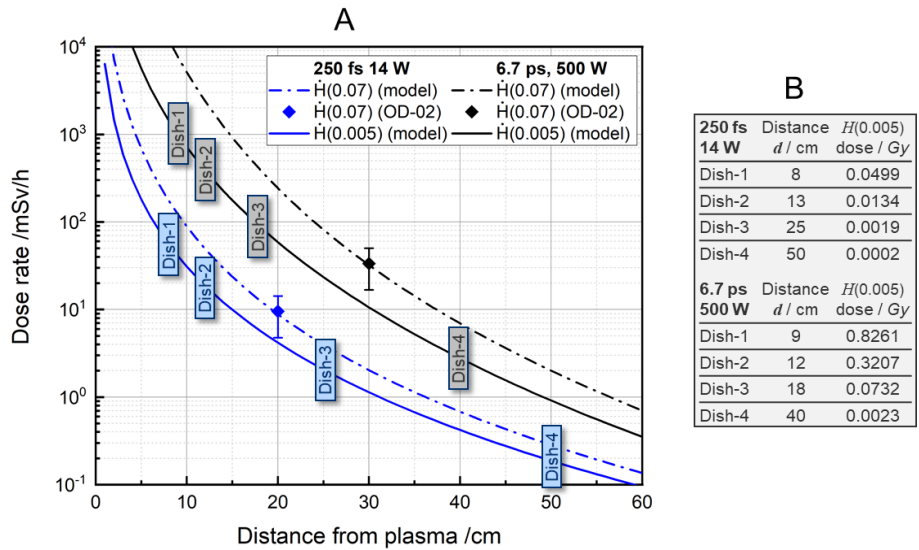

Figure 3. A) Dose rates as a function of the distance from the laser-induced plasma. The diamonds are the $\dot{H}(0.07)$-dose rates measuerd with the OD-02, the dash-dotted lines the corresponding model calculations. The solid lines are the model calculations for the $\dot{H}(0.005)$-dose rates as chosen as relevant for the HeLa cell layers. The blue values represent the experiments with 250 fs laser pulses and an average power of 14 W, the black values correspond to the experiments with 6.7 ps laser pulses and an average power of 500 W. The rectangles indicate the four positions of the dishes in each setup. B) Resulting dose after the exposure in the experiments.

It is seen that dose rates decrease very strongly with increasing distance from the laser-induced plasma, being already a factor of 10 000 lower at the distance of 60 cm as compared to a about factor of 10 closer distance of 5 cm. Furthermore, the dose rate increases linearly with the average laser power and in addition increases with the pulse duration[3,22]. Therefore, the 6.7 ps, 500 W setup generated significantly higher dose rates than the 250 fs, 14 W setup, even if the irradiance of the latter was $8.7 \cdot 10^{13}$ W/cm², which is much higher than the $2.5 \cdot 10^{13}$ W/cm² applied with the 6.7 ps, 500 W setup. The maximum dose rate at the nearest dish was 60 mSv/h and 1 000 mSv/h for the 250-fs setup and the 6.7-ps setup, respectively. It is noted the current legal limit in Germany for safe operation of an ultrafast laser processing system for the local dose rate is 10 µSv/h at 10 cm of the touchable surface[25]. This means that, except for Dish-4 at 250 fs, the dose rates used in our experiments were at least a factor of 4 000 higher than this threshold.

Determination of the dose

During the irradiation of the cells, the dose rates at the location of each cell dish integrate to a resulting dose, which is mainly relevant for potential hazard. This resulting dose was calculated by multiplying the value of the dose rate, calculated with the model, with the experimentally applied duration of the irradiation. Figure 3B summarizes the resulting $H$(0.005)-doses for each dish. For comparison with literature values (see below), the doses are given in Gy, where 1 Gy = 1 Sv in our case[26].

Immunofluorescence staining

Following the exposure to X-rays, the HeLa cells were allowed to recover for 1 h at 37°C in a humidified incubator with 5% $CO_2$. The non-exposed control cells were treated similarly. Subsequently, the samples were fixed for 15 min at room temperature with 4% (w/v) paraformaldehyde. After washing with phosphate buffer saline (PBS), the cells were permeabilized for 10 min at room temperature with 0.3% (v/v) Triton X-100 in PBS. Blocking was performed with 5% (v/v) goat serum (Invitrogen) in PBS containing 0.1% (v/v) Tween-20. Fixed samples were incubated overnight at 4°C with primary antibodies diluted in blocking buffer. A phospho-Histone H2A.X (Ser139) antibody (Cell Signaling, # 9718) was used to selectively recognize the γH2AX modification, while a histone H3 antibody (Santa Cruz, sc-517576) was used to counterstain the whole chromatin. Following three washing steps with PBS, the cells were finally incubated with Alexa-Fluor®-(488, 555) labelled secondary antibodies in a blocking buffer for 1 h at room temperature. The processed samples were stored and protected from light at 4°C. Imaging was performed within the next 48 h.

Fluorescence imaging and quantification of the γH2AX signal

Epifluorescence 16-bit TIFF images of the stained cells were taken on an EVOS M5000 microscope (Thermo Fischer Scientific) using a 20x objective. For quantification, 4 to 5 fields of view were arbitrary chosen per condition and imaged with identical acquisition settings among conditions that were to be directly compared. The CellProfiler[27] image analysis software was subsequently used to quantify the mean fluorescence intensity of the γH2AX signal at the single cell level. To this end, cell segmentation was performed based on the H3 co-staining. The resulting values of the mean fluorescence intensity of γH2AX were extracted for each nucleus and the values were compiled in Excel for further analysis. Statistical testing was performed by one-way ANOVA and Dunnette's post-test using GraphPad Prism vs 7.03. Please note that we chose to quantify the mean fluorescence intensity instead of γH2AX foci count, as done in other studies, since, due to our experimental setup, the cells had to be cultured on PS-dishes. This material is not compatible with the high-resolution imaging that is required to distinguish individual γH2AX foci.

## **Results**

Analysis of the DSBs response in HeLa cell cultures to repetitive, ultra-short X ray pulses

The human epithelial cell line HeLa was used to investigate the DSBs response upon exposure to the X-ray emission, which is generated during ultrafast laser processing of steel. These cells are frequently employed to study in vitro how exposure to X-irradiation activates the DNA damage response pathway[28]. To this end, cell monolayers were exposed to X-rays produced with two different laser systems, whereby the emission dose and pulse length was varied. Figure 4 shows the DSBs response, which was monitored by immunofluorescence microscopy and quantified at the single cell level on the basis of the γH2AX signal intensity.

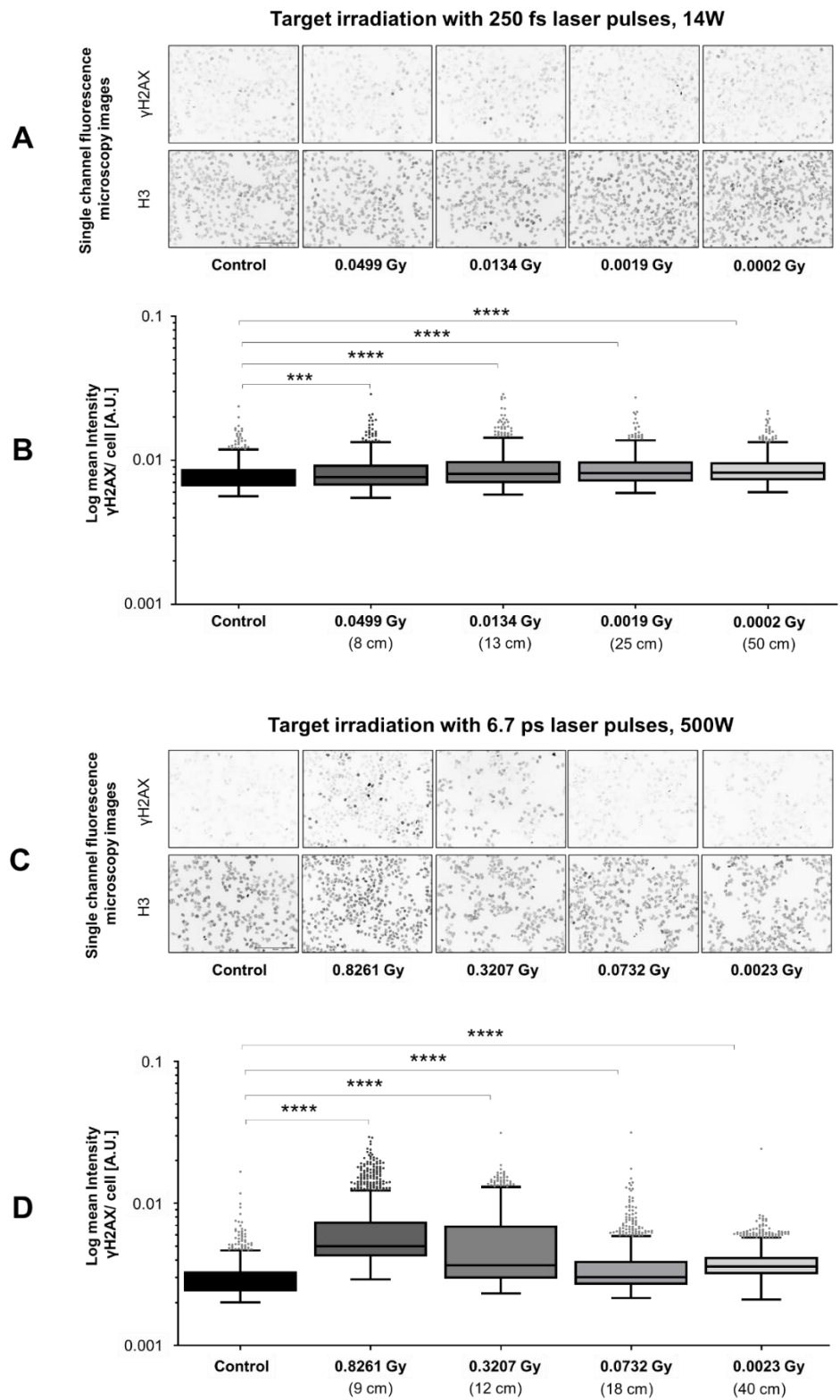

*Figure 4. DSBs response in HeLa cells induced by repetitive, ultra-short X-ray pulses resulting from target irradiation with the 250 fs and 14 W setup (A and B) and the 6.7 ps and 500 W setup (C and D). Representative single channel fluorescence microscopy images are shown in grey scale. All images of the γH2AX (DSBs marker) and the H3 (single cell nuclei) signal, respectively, which are shown within one panel are acquired and displayed with the same settings. Scalebar is 200 μm. (B and D) The box plots show the quantification of the mean γH2AX fluorescence signal per cell, in the experiment representatively shown in A and C. Each dot represents a single cell identified on the basis of the H3 staining. Center lines show the medians; box limits indicate the 25th and 75th percentiles as determined by GraphPad Prism 7 software; whiskers extend 1.5 times the interquartile range from the 25th and 75th percentiles, outliers are represented by dots. Each column collects data from circa 1000 single cells. Statistical significance was assessed by one-way ANOVA and Dunnette's post-test (N > 1000 cells/condition). \*\*\*p<0.001, \*\*\*\*p<0.0001.*

Single cell nuclei were identified on the basis of the H3 staining. The doses are given in Gy, where 1 Gy = 1 Sv[26]. For the measurements, 4 dishes with HeLa cells monolayers were placed at increasing distances from the processed material as indicated in Figure 4. One additional dish, outside of the laser enclosure, served as the non-exposed control.

With the 250 fs and 14 W setup, where the maximum dose was 0.049 Gy, the cells from the exposed dishes showed statistically significant but subtle differences in the γH2AX signal intensity relative to sham control (Figure 4A and 4B). By contrast, with the 6.7 ps and 500 W setup, there was a clearly visible and distance-dependent increase of the γH2AX signal in the exposed cells by comparison to the sham control (Figure 4C and D). This increase was strongest for dishes positioned closest to the processed material, where the radiation dose was 0.8261 Gy. By contrast to the γH2AX signal, no gross changes were observed in overall nuclear morphology under our experimental conditions, indicating that with our setup we detect an early stage in the molecular process of DNA damage response.

<u>HeLa cells show a linear DSBs response to ultra-short pulses of X-radiation</u>

Figure 5 shows the DSBs response plotted as a function of the applied $H(0.005)$-dose. This data representation mode was selected to facilitate comparison between our experimental setup utilizing X-ray pulses and prior studies employing X-ray tubes with continuous beams[29–31]. To extract the DSB response specifically triggered by exposure, the median γH2AX fluorescent signal shown in Figure 4 was normalized to the signal of the control dishes.

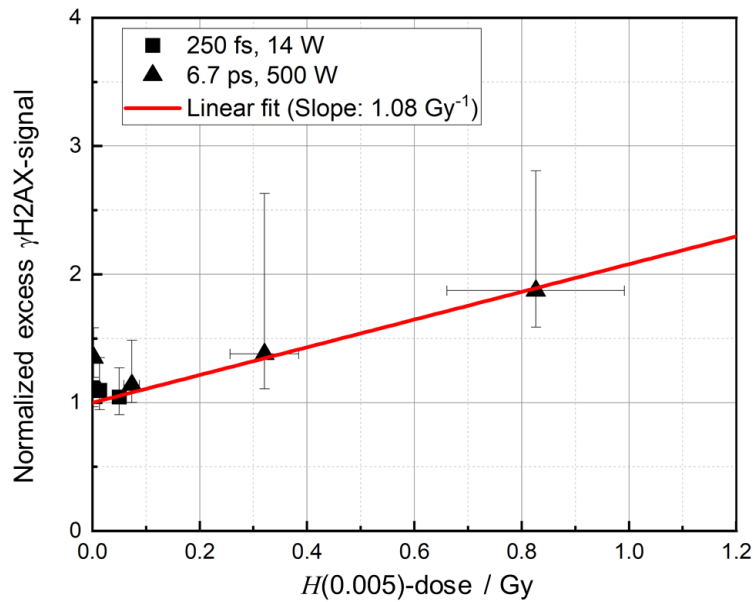

*Figure 5. Quantification of the normalized excess γH2AX-signal in the exposed cells relative the sham control #and displayed as a function of the H(0.005) dose. The data points refer to the median values of the results shown in Figure 4. Squares and triangles indicate the data obtained with the 250 fs and 14 W setup and the 6.7 ps and 500 W setup, respectively. The red line is a linear fit ($R^2 \approx 0.99$) to the datapoints, forced to 1 for H(0.005) = 0 Gy. The error bars represent the 25% as well as the 75% quantile of the intensities of the fluorescence signal (Figure 4).*

The error bars represent the 25% as well as the 75% quantile of the intensities of the fluorescence signal (Figure 4). The red dashed line is a linear fit to the data points, forced to equal to 1 for the dose of $H(0.005) = 0$. The slope of the linear fit is 1.08 Gy$^{-1}$ and the coefficient of determination is $R^2 = 0.995$. Based on this analysis, it was concluded that the DSBs response increases linearly with the applied $H(0.005)$-dose.

## Discussion and conclusion

Materials processing with ultrafast lasers, delivering pulses with durations between a few hundred femtoseconds and a few picoseconds, enables precise processing of a wide range of materials. This capability holds significant interest for the industry, offering a cutting-edge approach to enhance manufacturing processes and enable the fabrication of intricate components with unparalleled accuracy[1]. Nonetheless, the generation of photons with energies above 5 keV during this process has sparked extensive debates on the radiation safety implications associated with the ultrafast laser processing.

In this study we investigated for the first time the molecular consequences of exposure to the repetitive, ultra-short X-ray pulses generated during laser processing of steel. To this end, we examined the DNA damage response of in-vitro exposed HeLa cell cultures, whereby conditions with 250 fs and 6.7 ps laser pulses were compared. The photon energy of the maximum of the characteristic bremsstrahlung emission from our X-ray source at the position of the cells was in the range of 3 to 6 keV. The maximum dose applied to the cells was 0.826 Gy using 6.7 ps laser pulses at 500 W. As documented in Figure 5, a linear increase of the DSBs response with increasing dose was found. This finding aligns with similar trends observed in other studies using X-ray tubes with continuous emission[29–31], where a linear correlation was reported between the DSB response and the dose of X-ray exposure, with slopes ranging between 2 $Gy^{-1}$ and 5 $Gy^{-1}$. However, by contrast to the continuous radiation of the X-ray tubes used in these reports, in our study, the radiation results from plasma, which was produced with ultra-short laser pulses with a duration of either 250 fs or 6.7 ps. The produced X-ray pulses were repetitive and had a similar ultra-short pulse duration as the producing laser pulses, i.e. in the order of magnitude of picoseconds. This results in very high peaks in the photon flux of the X-ray photons. Further, as shown in Figure 2, the energy distribution of the photons is different, and its maximum is in the soft X-ray energy region at around 4 keV. In spite of these differences, our data indicate that under our experimental conditions there is no additional increase of the DSBs response when the exposure of the cells occurs with ultra-short X-ray pulses, i.e., there was no evidence for additional hazard for ultra-short X-ray pulses. This is supported by the fact that at relatively comparable doses of 0.05 Gy vs 0.07 Gy, a factor of 27 in the duration of the laser pulses from 250 fs to 6.7 ps did not cause an increase in the measured DNA damage response.

Considering the complexity of DNA damage response pathways and their variability among cellular systems[32], future investigations should delve more extensively into these molecular findings, for instance by incorporating skin-on-a-chip models. This exploration will contribute to a deeper understanding of the implications of ultrafast laser processing on biological systems and guide safety considerations in both industrial and biomedical applications.

**Data availability**
The data that supports the findings of the study are included in the main text of the manuscript. Raw data can be obtained from the corresponding authors upon request.

**Contributions**
RW, JH, CL conceived the research and designed experiments with contribution from ME. RW and CL supervised the research. JH and CL conducted experiments. JH, CL, RW analyzed the data. All authors discussed the data. JH, CL and RW wrote the draft article and prepared the figures. CL and TG acquired funding. All authors were involved in preparing the final manuscript.

**Acknowledgements**
This work was supported by the Baden-Wuerttemberg Ministry of Science, Research and Arts by a grant to CL and providing infrastructure to JH, RW, and TG. Part of the work was funded by the Deutsche Forschungsgemeinschaft (DFG, German Research Foundation) in the frame of INST 41/1031 1 FUGG and the project 491192473.

**Ethics declaration**
The authors declare no competing interests.